\documentclass[floatfix,reprint,amsmath,amssymb,aps,superscriptaddress]{revtex4-2}
\pdfoutput=1

\usepackage{algorithmic}
\usepackage{graphicx}
\usepackage{caption, subcaption}
\usepackage{textcomp}
\usepackage[dvipsnames]{xcolor}
\def\BibTeX{{\rm B\kern-.05em{\sc i\kern-.025em b}\kern-.08em
    T\kern-.1667em\lower.7ex\hbox{E}\kern-.125emX}}

\usepackage{braket}
\usepackage{float}

\usepackage{hyperref}
\hypersetup{
  breaklinks=true,
  colorlinks=true,
  allcolors=BlueViolet,
}
\usepackage[capitalise]{cleveref}

\Crefname{figure}{Fig.}{Figs.}
\crefname{equation}{}{}

\usepackage[ruled]{algorithm2e}
\usepackage{tikz}
\usetikzlibrary{quantikz2}

\begin{document}

\title{Average circuit eigenvalue sampling on NISQ devices}

\author{Emilio Peláez Cisneros}
\affiliation{University of Chicago, Chicago, IL 60637, USA}
\affiliation{Infleqtion, Chicago, IL 60604, USA}

\author{Victory Omole}
\affiliation{Infleqtion, Chicago, IL 60604, USA}

\author{Pranav Gokhale}
\affiliation{Infleqtion, Chicago, IL 60604, USA}

\author{Rich Rines}
\affiliation{Infleqtion, Chicago, IL 60604, USA}

\author{Kaitlin N.~Smith}
\affiliation{Infleqtion, Chicago, IL 60604, USA}

\author{Michael A.~Perlin}
\affiliation{Infleqtion, Chicago, IL 60604, USA}

\author{Akel Hashim}
\affiliation{University of California at Berkeley, CA 94720, USA}`  

\date{\today}

\begin{abstract}
Average circuit eigenvalue sampling (ACES) was introduced by Flammia in Ref.~\cite{Flammia2021AveragedCE} as a protocol to characterize the Pauli error channels of individual gates across the device simultaneously. The original paper posed using ACES to characterize near-term devices as an open problem. This work advances in this direction by presenting a full implementation of ACES for real devices and deploying it to Superstaq \cite{superstaq}, along with a device-tailored resource estimation obtained through simulations and experiments. Our simulations show that ACES is able to estimate one- and two-qubit non-uniform Pauli error channels to an average eigenvalue absolute error of under $0.003$ and total variation distance of under 0.001 between simulated and reconstructed probability distributions over Pauli errors
with $10^5$ shots per circuit using 5 circuits of depth 14. The question of estimating general error channels through twirling techniques in real devices remains open, as it is dependent on a device's native gates, but simulations with the Clifford set show results in agreement with reported hardware data. Experimental results on IBM's Algiers and Osaka devices are presented, where we characterize their error channels as Pauli channels without twirling.
\end{abstract}

\maketitle

\section{Background}
Learning a quantum noise channel is essential to useful quantum computation. Usually, errors are classified as coherent or incoherent. We are concerned with the latter. If an incoherent noise channel is characterized successfully, it can be used to perform error mitigation of successive experiments done in the characterized device \cite{cao2022nisq}. Thanks to this and other use cases, many noise-learning techniques have been developed, many of which draw inspiration from randomized benchmarking \cite{emerson2005scalable, Knill_2008}. Average circuit eigenvalue sampling (ACES) is a quantum noise characterization protocol introduced in \cite{Flammia2021AveragedCE}. ACES estimate the parameters it is trying to characterize in parallel, therefore allowing it to compute a large number of parameters with a  low number of circuits in comparison.

In this paper, we first reintroduce the ACES protocol and necessary background to understand it. Then, we outline a step-by-step procedure of ACES (which we have deployed to Superstaq \cite{superstaq}), simulate its performance against a Pauli error channel, and run it on IBM quantum hardware. We find that ACES reconstructs the Pauli error channels with a total variation distance of under 0.001 between simulated and reconstructed probability distributions over Pauli errors using $10^5$ measurements per circuit and below 0.004 using $10^4$ measurements per circuit in simulation. 
On hardware, we learn two-qubit error rates consistent with IBM's reported values to a relative error of 12\% and 23\% in the best cases, but over an order of magnitude away in the worst case (see Section \ref{sec:discussion}). To the best of our knowledge, this paper presents the first results of running ACES on publicly available hardware.

\subsection{Twirling and eigenvalues}

General quantum channels $\mathcal{E}$ act on density operators $\rho$ as $\mathcal{E}(\rho) = \sum_i K_i \rho K_i^\dagger$, where $K_i$ are Kraus operators. A Pauli noise channel over $n$ qubits is a quantum channel of the form $\rho \rightarrow \sum_{P_a \in \mathbb{P}^n} p_a P_a \rho P_a^\dagger$, where $\mathbb{P}^n$ is the group of $n$-qubit operators. Essentially, this channel applies the Pauli operator $P_a$ to $\rho$ with probability $p_a$. Given a general quantum channel $\mathcal{E}$, we can twirl it by the $n$-dimensional Pauli group to make it a Pauli channel $\mathcal{E}^P$ \cite{Nielsen_2002}. This operation looks as follows.

\begin{equation}
    \mathcal{E}^P(\rho) = \frac{1}{4^n} \sum_{P_a \in \mathbb{P}^n} P_a^\dagger \mathcal{E}(P_a \rho P_a^\dagger)P_a
\end{equation}

At the circuit level, this can be done by choosing a random Pauli and placing it before and after the original channel $\mathcal{E}$. This random process is repeated multiple times and the collection of twirled circuits over the Pauli group constitute $\mathcal{E}^P$. This process is usually referred to as Pauli twirling.

Given a channel $\mathcal{E}$ with Kraus operators $K_j = \sum_a \nu_{j, a}P_a$, the twirled channel $\mathcal{E}^P$ has Pauli error rates $p_a = \sum_j |\nu_{j, a}|^2$ \cite{Flammia2021AveragedCE}. We can consider the eigenvectors of this channel to be the Pauli operators, such that $\mathcal{E}^P(P_b) = \lambda_b P_b$ for
\begin{equation}\label{eq:rates-to-eig}
    \lambda_b = \sum_b (-1)^{\langle a, b \rangle} p_a.
\end{equation}
Inverting this equation we get
\begin{equation}\label{eq:eig-to-rates}
    p_a = \frac{1}{2^n}\sum_b (-1)^{\langle a, b \rangle} \lambda_b.
\end{equation}
With this, we have a way to infer the Pauli error rates of a channel from the Pauli eigenvalues of the same channel.

\begin{figure}[H]
    \centering
    \begin{quantikz}
        &\gate{\tilde{\mathcal{G}}}&
    \end{quantikz} $\rightarrow$
    \begin{quantikz}
        &\gate{P_a}&\gate{\tilde{\mathcal{G}}}&\gate{P_{a'}} &
    \end{quantikz}
    \caption{$\mathcal{G}$-twisted Pauli twirl of a Pauli $P_a$}
    \label{fig:g-twirl}
\end{figure}
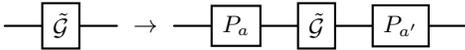

Another twirling technique, used in \cite{Flammia2021AveragedCE}, is the $\mathcal{G}$-twisted Pauli twirl. Consider a Clifford gate $\mathcal{G}$ and its noisy implementation $\tilde{\mathcal{G}}$. We can think of the noisy implementation as the composition of the ideal channel and a noisy channel as $\tilde{\mathcal{G}} = \mathcal{GE}$, where we have that $\mathcal{E} = \mathcal{G}^\dagger \tilde{\mathcal{G}}$ is the noisy channel. The $\mathcal{G}$-twisted Pauli twirl is then defined as
\begin{equation}
    \tilde{\mathcal{G}}^{\mathcal{G}P}(\rho) = \frac{1}{4^n}\sum_a P_{a'}^\dagger \tilde{\mathcal{G}}(P_a \rho P_a^\dagger) P_{a'},
\end{equation}
where $P_{a'} = \mathcal{G}(P_a) = \mathcal{G}P_a \mathcal{G}^\dagger$ is the conjugation of $P_a$ by $\mathcal{G}$. We have that this is equal to $\mathcal{G}(\mathcal{E}^P(\rho))$, which shows that this twirling isolates the noise around the noisy gate $\tilde{\mathcal{G}}$. The equality can be shown by expanding the definitions of each channel as follows.
\begin{align}
    \mathcal{G}(\mathcal{E}^P(\rho)) &= \frac{1}{4^n} \sum_a \mathcal{G} P_{a}^\dagger \mathcal{E}(P_a \rho P_a^\dagger) P_a \mathcal{G}^\dagger \\
    &= \frac{1}{4^n} \sum_a P_{a'}^\dagger \mathcal{GE}(P_a \rho P_a^\dagger) \mathcal{G}^\dagger P_{a'} \\
    &= \frac{1}{4^n} \sum_a P_{a'}^\dagger \mathcal{G}(\mathcal{E}(P_a\rho P_a^\dagger)) P_{a'} \\
    &= \frac{1}{4^n} \sum_a P_{a'}^\dagger \tilde{\mathcal{G}}(P_a \rho P_a^\dagger) P_{a'} \\ 
    &= \tilde{\mathcal{G}}^{\mathcal{G}P}(\rho)
\end{align}
Analogous to the normal Pauli twirl, this consists of randomly choosing a Pauli $P_a$ and appending it before $\tilde{\mathcal{G}}$. Then, $P_{a'}$ can computed efficiently and appended after $\tilde{\mathcal{G}}$. This is illustrated in Fig. \ref{fig:g-twirl}. The twirl is then a collection of circuits obtained by performing this random process multiple times over the Pauli group.

Analogous to how eigenvalues are defined when we were considering the Pauli eigenvalues of a Pauli twirled channel, we can define the generalized eigenvalues of a $\mathcal{G}$-twisted error channel. Given a channel $\tilde{\mathcal{G}}^{\mathcal{G}P}$, we have that $\tilde{\mathcal{G}}^{\mathcal{G}P}(\rho) = \lambda \mathcal{G}_0 (\rho)$, where $\lambda$ is the generalized eigenvalue with respect to the generalized eigenvector $\mathcal{G}_0(\rho)$. Setting $\mathcal{G}_0 = \mathcal{G}$ gives the Paulis as generalized eigenvectors with eigenvalues being the Pauli eigenvalues of $\mathcal{E}^P$ \cite{Flammia2021AveragedCE}. In our case, we can think of the channels we are going to study as Clifford 
circuits. Consider a circuit $\mathcal{C}$ of depth $T$ is made up of a sequence of Clifford gates $\{\mathcal{G}_i, \ldots, \mathcal{G}_T\}$, such that $\mathcal{C} = \mathcal{G}_T\cdots\mathcal{G}_1$. Applying the first gate to some input $P_{a_1}$, we have that $\tilde{\mathcal{G}}_1(P_{a_1}) = \pm \lambda_{1,a_{1}} P_{a_2}$. Applying each gate in succession, we can find a relation between circuit and gate eigenvalues as follows.
\begin{align}
    \tilde{\mathcal{C}}(P_{a_{1}}) = \pm \prod_{k=1}^T \lambda_{k,a_{k}} \mathcal{C}(P_a{1})
\end{align}
The sign in this expression and the output Pauli can be efficiently classically computed since we are only considering Clifford circuits. Defining a circuit eigenvalue as $\Lambda_{\mathcal{C}, a_{1}}$, we get that
\begin{align}\label{eq:circuit-and-gate-eigenvalues}
    \Lambda_{\mathcal{C}, a_{1}} = \prod_{k=1}^T \lambda_{k, a_{k}}.
\end{align}
This gives us a relationship between circuit and gate eigenvalues, which we can use to get the latter from the former in the model we are going to define later on.

\subsection{Circuit averaging}

Given a noisy Clifford circuit $\mathcal{C}$, we can create a new circuit with the same overall unitary through randomized compiling \cite{Wallman_2016, hashim2021randomized} by applying a $\mathcal{G}$-twisted Pauli twirl to each gate in the circuit. That is, applying a random Pauli $P_a$ before every Clifford gate $\mathcal{G}$, and the corresponding $P_{a'} = \mathcal{G} P_a \mathcal{G}^\dagger$ after $\mathcal{G}$. Doing this multiple times, we can create an ensemble of circuits $\mathcal{C}^P$.
This way, we effectively average the noise over the Pauli group and we can study it as a Pauli channel, which we aim to estimate. Therefore, given an original circuit $\mathcal{C} = \mathcal{G}_T \cdots \mathcal{G}_1$, its averaged noisy implementation will be $\tilde{\mathcal{C}}^P = \widetilde{\mathcal{G}_T}^{\mathcal{G}_T P}\cdots \widetilde{\mathcal{G}_1}^{\mathcal{G}_1 P}$. 

By circuit averaging, we mean the averaging over random choices of Paulis inserted as detailed above. This
was used when simulating ACES, but when running experiments on IBM devices, we did not use this technique because, with existing hardware, it is cheaper to run more shots of the same circuit than the same number of shots distributed over multiple circuits. 
Therefore, the hardware results presented in this paper approximate the device error channel as a stochastic Pauli error channel without using its averaged implementation. 

\subsection{Eigenvalue sampling}
Consider an ideal Clifford circuit $\mathcal{C}$. Given its noisy averaged implementation $\tilde{\mathcal{C}}^P$, we want to estimate its circuit eigenvalue with respect to some Pauli operator $P_{a_1}$. Recall from earlier that this is denoted $\Lambda_{\mathcal{C}, a_1}$ and given by $\tilde{\mathcal{C}}^P(P_{a_1}) = \Lambda_{\mathcal{C}, a_1}  \mathcal{C}(P_{a_1}) = \pm \Lambda_{\mathcal{C}, a_1} P_{a_2}$. It follows that we can estimate the value for this circuit eigenvalue by using that
\begin{equation}\label{eq:circ-eig-trace}
    \Lambda_{C,a_1} = \frac{1}{2^n} \mathrm{Tr} \left( P_{a_2} \tilde{\mathcal{C}}^P(P_{a_1}) \right).
\end{equation}

Eigenvalue sampling is the process of estimating $\Lambda_{C,a_1}$ through the expression above. In the right hand side of this equation, we have an expectation value. This suggests that taking repeated measurements of $\tilde{\mathcal{C}}^P(P_{a_1})$ in the $P_{a_2}$ basis will give us an estimate of $\Lambda_{C,a_1}$. The exact experimental procedure we use to do this is detailed in Section \ref{sec:estimating} below.

\section{Implementing ACES in hardware}\label{sec:implementing}

The goal of ACES is to characterize the Pauli error channel of each gate in a Clifford gate set $\mathbf{G}$. For each $\mathcal{G} \in \mathbf{G}$, we denote by $\mathrm{size}(\mathcal{G})$ the number of qubits the gate acts on. For each single-qubit gate, we want to solve for three parameters: $p_x, p_y, p_z$. For every two-qubit gate, we want to solve for $15$ parameters, i.e., every pair of Paulis $p_{ij}$ except for $II$. To do this, we seek to learn a collection of circuit eigenvalues from which we can infer the individual gate eigenvalues. 

Consider a circuit and the relationship between its circuit and gate eigenvalues as done in \eqref{eq:circuit-and-gate-eigenvalues}. There, we show that the circuit eigenvalue is a product of the eigenvalues of the gates that make up the circuit. We want to exploit this relationship to find the gate eigenvalues given the circuit eigenvalue. We can take the logarithm of both sides of \eqref{eq:circuit-and-gate-eigenvalues} to convert our problem into a set of linear equations, as follows:
\begin{align}
    \Lambda_{\mathcal{C}, a_{1}} &= \lambda_{1, a_1} \cdot \ldots \cdot \lambda_{T, a_T} \\
    \ln\left(\Lambda_{\mathcal{C}, a_{1}}\right) &= \ln\left( \lambda_{1, a_1} \cdot \ldots \cdot \lambda_{T, a_T} \right) \\
    \ln\left(\Lambda_{\mathcal{C}, a_{1}}\right) &= \ln\left(\lambda_{1, a_1}\right) + \cdots + \ln\left(\lambda_{T, a_T}\right) \label{eq:ln-circ-gate-eig}
\end{align}
In \eqref{eq:ln-circ-gate-eig}, we have a linear relationship between the logarithm of the circuit and gate eigenvalues. Define new variables as $b_{\mathcal{C}, a_k} = \ln\left( \Lambda_{\mathcal{C}, a_k} \right)$ and $x_{k, a_k} = \ln\left( \lambda_{k, a_k} \right)$ and consider a collection of circuits $\{\mathcal{C}_0, \mathcal{C}_1, \ldots, \mathcal{C}_N\}$ instead of a single circuit. Relabel the composite indices of our new variables as $\mu = (\mathcal{C}_i, a_{i_k})$ and $\upsilon = (k, a_k)$ for each circuit $\mathcal{C}_i$. Then, given a design matrix $A$ that contains data about the internal structure of the circuits that we are using to characterize our model, we get the following relation.
\begin{equation}\label{eq:linear-model-eq}
    \sum_{\upsilon} A_{\mu, \upsilon} x_{\upsilon} = b_\mu
\end{equation}
Thus, our problem is reduced to solving the matrix-vector equation $A\textbf{x} = \mathbf{b}$ to get an estimation of $\textbf{x}$. To get a good estimation, given that $A$ has shape $M\times N$, we need $A$ to have rank $N$. We can recover the gate eigenvalues by using the definition of the variables $x_{k, a_k}$. Finally, the Pauli error rates can be obtained through the relation in \eqref{eq:eig-to-rates}. Thus, at the end of the protocol, we get an estimation of the error channel as a Pauli channel.

\subsection{Protocol}

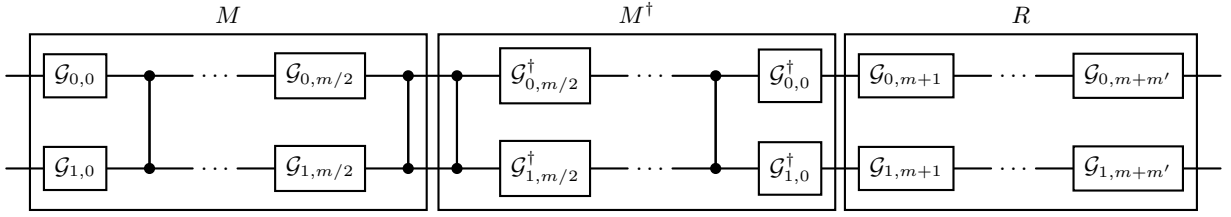
\begin{figure*}
    \centering
    \begin{quantikz}
        & \gate{\mathcal{G}_{0,0}}\gategroup[2,steps=5,style={inner sep=2pt}]{$M$} & \ctrl{1} & \ \ldots\ &  \gate{\mathcal{G}_{0,m/2}} & \ctrl{1} & \ctrl{1}\gategroup[2,steps=5,style={inner sep=2pt}]{$M^\dagger$} & \gate{\mathcal{G}_{0,m/2}^\dagger} & \ \ldots\ & \ctrl{1} & \gate{\mathcal{G}_{0,0}^\dagger} & \gate{\mathcal{G}_{0,m+1}}\gategroup[2,steps=3,style={inner sep=2pt}]{$R$} & \ \ldots\ & \gate{\mathcal{G}_{0,m+m'}} & \\
        & \gate{\mathcal{G}_{1,0}} & \ctrl{-1} & \ \ldots\ & \gate{\mathcal{G}_{1,m/2}} & \ctrl{-1} & \ctrl{-1} & \gate{\mathcal{G}_{1,m/2}^\dagger} & \ \ldots\ & \ctrl{-1} & \gate{\mathcal{G}_{1,0}^\dagger} & \gate{\mathcal{G}_{1,m+1}} & \ \ldots\ & \gate{\mathcal{G}_{1,m+m'}} &
    \end{quantikz} 
    \caption{Structure of random circuits $\mathcal{C}_i$.}
    \label{fig:random-circuits}
\end{figure*}

We first create a collection of $N$ random circuits
$\{\mathcal{C}_1, \mathcal{C}_2, \ldots, \mathcal{C}_N\}$.
Each $\mathcal{C}_i$ will have two parts in its structure: a mirror section and a random section. The mirror section will consist of $m/2$ moments of alternating layers of one-qubit and two-qubit gates randomly sampled from $\mathbf{G}$, followed by the inverse of these moments. Therefore, the overall effect of this section is the identity. Then, $m'$ random moments with one-qubit and two-qubit gates in no particular order are appended, giving a total circuit depth of $m + m'$. This structure is illustrated in Fig. \ref{fig:random-circuits}.

Other circuit structures could be considered, but we found that this particular one yielded results with low total variation distance between simulated and reconstructed Pauli channels and low absolute error in the gate eigenvalues, as presented in the results section below.

\subsubsection{Estimating circuit eigenvalues}\label{sec:estimating}

For each circuit $\mathcal{C}_i$, we have to determine a set $\mathcal{P}_{C_i}$ of Pauli operators for which we are going to estimate their circuit eigenvalues. There are many criteria that can be taken to construct each $\mathcal{P}_{C_i}$, here we choose Pauli strings of weight one or two that remain of weight one or two after going through $\mathcal{C}_i$. This is done to try to minimize the number of qubits that have to be measured.
After we determine this set of Paulis, we seek to estimate each $\Lambda_{C_i, j}$ for $P_j \in \mathcal{P}_{C_i}$. 

From \eqref{eq:circ-eig-trace}, we know that we want to measure the expectation value of $\tilde{\mathcal{C}}_i(P_j)$ in the $\mathcal{C}_i(P_j)$ basis (which we know is a Pauli basis since it is a Clifford circuit applied to a Pauli operator). However, the density matrix corresponding to $P_j$ is not a physical state that can be prepared, so a direct implementation is not possible. 

Therefore, we use the \textit{difference trick} to estimate each $\Lambda_{C_i, j}$. For this, consider the sets of positive and negative eigenvectors, $\Psi_{j}^+$ and $\Psi_{j}^-$, of $P_j$ with eigenvalues $\pm 1$. With these eigenvectors, we can compute $\Lambda_{C_i,j}$ through
\begin{align}\label{eq:difference-trick}
\begin{split}    
    \Lambda_{C_i, j} = \frac{1}{2^n}\Biggl[ &\sum_{\psi^+ \in \Psi_{j}^+} \mathrm{Tr}\left( P_{j'} \tilde{C}_i^P(\psi^+) \right) \\ &- \sum_{\psi^- \in \Psi_{j}^-} \mathrm{Tr}\left( P_{j'} \tilde{C}_i^P(\psi^-) \right) \Biggr],
\end{split}
\end{align}
where each $\psi^+ \in \Psi_{j}^+$ and $\psi^- \in \Psi_{j}^-$ is an easy to prepare density matrix, as we explain below.

Consider the eigenstates of each of the Pauli matrices. For the $X$ operator, the eigenvectors are $|+\rangle$ and $|-\rangle$, prepared by $H|0\rangle$ and $HZ|0\rangle$, respectively. The $Y$ operator has eigenvectors $|-i\rangle$ and $|-i\rangle$, prepared by $SH|0\rangle$ and $S^{\dagger}H|0\rangle$. The eigenvectors of the $Z$ operator are $|0\rangle$ and $|1\rangle$, which are trivially initialized. For some Pauli string $P_j \in \mathbb{P}^n$, that is, a sequence of length $n$ of $X, Y, Z, I$, we prepare an eigenstate with eigenvalue $+1$ or $-1$ by going through the Pauli in each qubit and preparing one of its eigenvectors. For the qubits in which $P_j$ has an $I$, we simply initialize them to the $|0\rangle$ state which has positive eigenvalue. When preparing each individual eigenvector, we have to keep track of the corresponding eigenvalue ($\pm 1$). At the end, if we used an even number of negative eigenvalues, the overall eigenvector will have eigenvalue $+1$. If we used an odd number of negative eigenvectors, the overall eigenvector will have eigenvalue $-1$.

Using this eigenvector preparation scheme we can obtain an estimation $\hat \Lambda_{\mathcal{C}_i, j}$ of the circuit eigenvalue $\Lambda_{\mathcal{C}_i, j}$ by preparing each of the eigenvectors of $P_j$, computing their expectation value in the $\mathcal{C}_i(P_j)$ measurement basis, and plugging the results into \eqref{eq:difference-trick}. 

\subsubsection{Design matrix construction}

Once we have a set of circuits and Pauli strings to measure for each circuit, we can construct a design matrix $A$. The number of rows of this matrix will be the total number of circuit eigenvalues $\left(\sum_i |\mathcal{P}_{C_i}|\right)$, and the number of columns will be the number of gate eigenvalues $\left( \sum_{\mathcal{G} \in \mathbf{G}} (4^{\mathrm{size}(\mathcal{G})} - 1)\right)$. The design matrix is then constructed as detailed in Algorithm \ref{alg:design-matrix}. As mentioned earlier, we want the design matrix to have rank equal to at least its number of columns, which we now know the value of. In our experiment, the parameters to obtain a design matrix of valid rank were obtained through trial-and-error.

\begin{algorithm}
\caption{Construct design matrix}\label{alg:design-matrix}
\KwData{Collection of circuits $\{\mathcal{C}_0, \mathcal{C}_1, \ldots, \mathcal{C}_N\}$, each with a set of probes $\mathcal{P}_{\mathcal{C}_i}$.}
\KwResult{Design matrix $A$.}
$r \leftarrow$ \# of circuit eigenvalues $\sum_i |\mathcal{P}_{C_i}|$\;
$c \leftarrow$ \# of gate eigenvalues $\sum_{\mathcal{G} \in \mathbf{G}} (4^{\mathrm{size}(\mathcal{G})} - 1)$\;
$A \leftarrow$ all-zero $r \times c$ matrix\;
$i \leftarrow 0$\;
\For{$\mathcal{C}_i \in \{\mathcal{C}_0, \mathcal{C}_1, \ldots, \mathcal{C}_N\}$}{
   \For{$P_j \in \mathcal{P}_{C_i}$}{
       Initialize current state $S \leftarrow P_j$\;
       \For{$\mathcal{G}_k \in \mathcal{C}_i$}{
           Find index $\mu = (S, G_k)$\;
           Update matrix $A[i][\mu]$ += $1$\; 
           Update state $S \leftarrow G_k(S)$\;
       }
       $i \leftarrow i + 1$\;
   }   
}
\end{algorithm}

\subsubsection{Solving the model}

The last two steps give us a design matrix $A$ and a collection of circuit eigenvalues in a vector $\textbf{b}$. Finally, we solve for $\textbf{x}$ in the equation $A\textbf{x} = \textbf{b}$ via least-squares to get the gate eigenvalues. From these gate eigenvalues, we obtain the Pauli error rates of the noise channel. This is the final output of the protocol.

\subsection{Resource estimation}
We can estimate how many circuits are needed to get a full rank design matrix $\textbf{A}$ by looking at how many ``linearly independent" circuits (the same circuit with different Pauli operator eigenvalues being estimated count as different circuits for this purpose) are needed. By linearly independent circuits we mean that their expansion to a sequence of gates cannot be obtained by a linear combination of the expansion of a number of other circuits. In other words, we want that expanding the eigenvalues into the expression in \eqref{eq:ln-circ-gate-eig} gives a linearly independent equation with respect to the other circuits. Therefore, we want at least as many independent circuits as gate eigenvalues we have. For a gateset with $N$ qubits,  $G_1$ one-qubit gates, and $G_2$ two-qubit gates, along with nearest-neighbour linear connectivity, we would need
\begin{align}
    3 \cdot G_1 \cdot N + 15 \cdot G_2 \cdot (N-1)
\end{align}
circuits. Then, extracting each expectation value with small additive error $\varepsilon$ requires $\mathcal{O}(1 / \varepsilon^2)$ measurements. For each circuit, there are $2^{w}$ eigenvalues to estimate, where $w$ is the weight of the Pauli string to probe. In theory, $w$ is upper bounded by the number of qubits in the circuits being sampled. In our experiments, we choose Pauli strings with $w\le 2$.

Therefore, to estimate all the eigenvalues with fidelity $\varepsilon$, we will need $\mathcal{O}(N(G_1 + G_2)\varepsilon^{-2})$ total measurements. In the following section we will provide empirical estimations about the resources needed to estimate the model parameters. 

\section{Simulation and experimental results}

For the real-device experiment and simulations, we seek to characterize the Pauli error channels of six one-qubit Clifford gates and the two-qubit CZ gate. For the one-qubit gates, we consider them on each qubit separately and we assume connectivity between successive qubits for the CZ gate. In both simulation and hardware runs we used a mirror depth $m' = 4$ and a random depth $m = 6$, giving circuits of total depth $14$. $5$ circuits are used in both simulations and hardware. We deployed the protocol to the Superstaq platform \cite{superstaq} and used it to run both the simulations and experiments on IBM hardware.

\subsection{Simulations}

\begin{figure}
     \centering
     \includegraphics[scale=0.5]{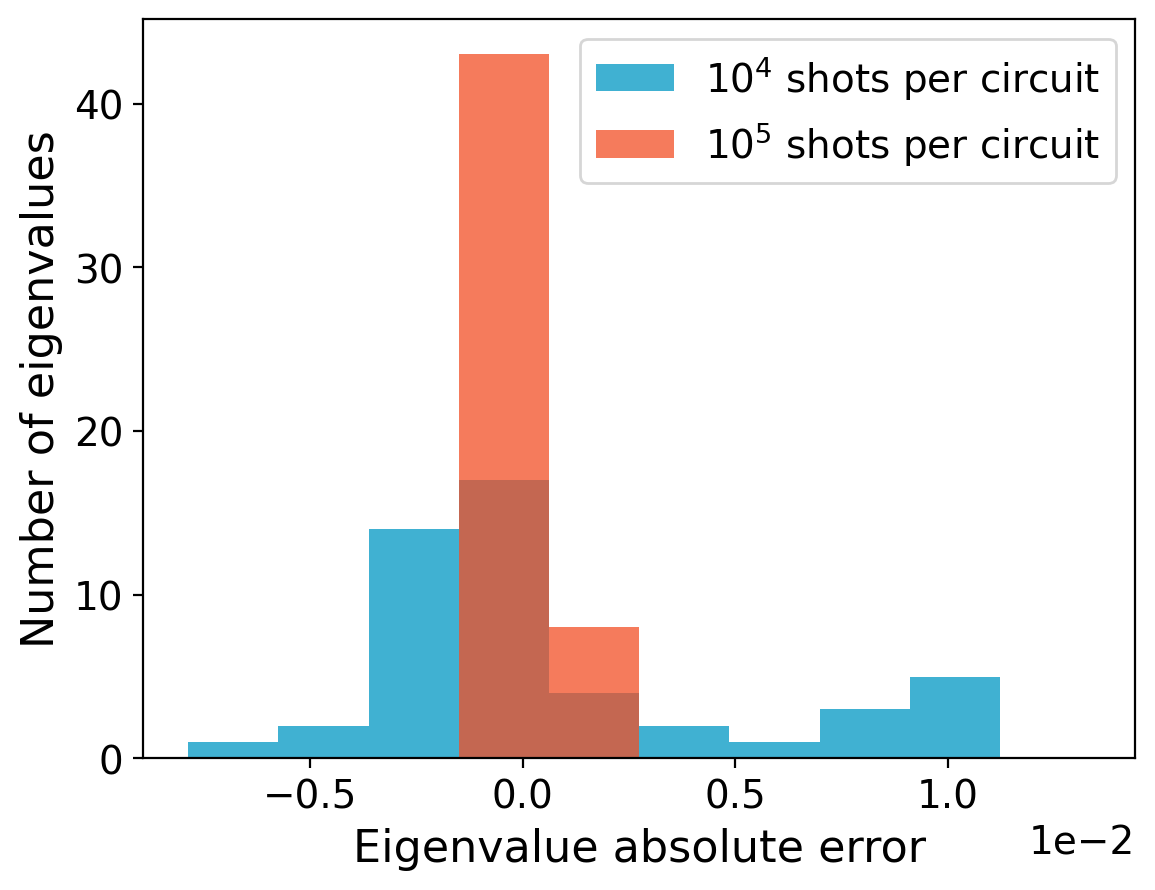}
     \caption{Absolute error of estimated gate eigenvalues in simulations with $10^4$ and $10^5$ shots per circuit.}
     \label{fig:circ-eig-abs-error}
\end{figure}

We first simulated the protocol detailed in the previous section to obtain an empirical estimation of how many shots were required to get accurate results on hardware. We ran the experiments using a simple stochastic noise model, i.e., we considered a purely Pauli noise channel. For each circuit, we use 10 $\mathcal{G}$-twisted Pauli twirls and divide the reported number of shots per circuit evenly between each twirled circuit. The absolute error between the estimated gate eigenvalues and the gate eigenvalues derived from the noise model used to simulate the circuits are presented in Figure \ref{fig:circ-eig-abs-error}.

\begin{figure}
    \centering
    \includegraphics[scale=0.5]{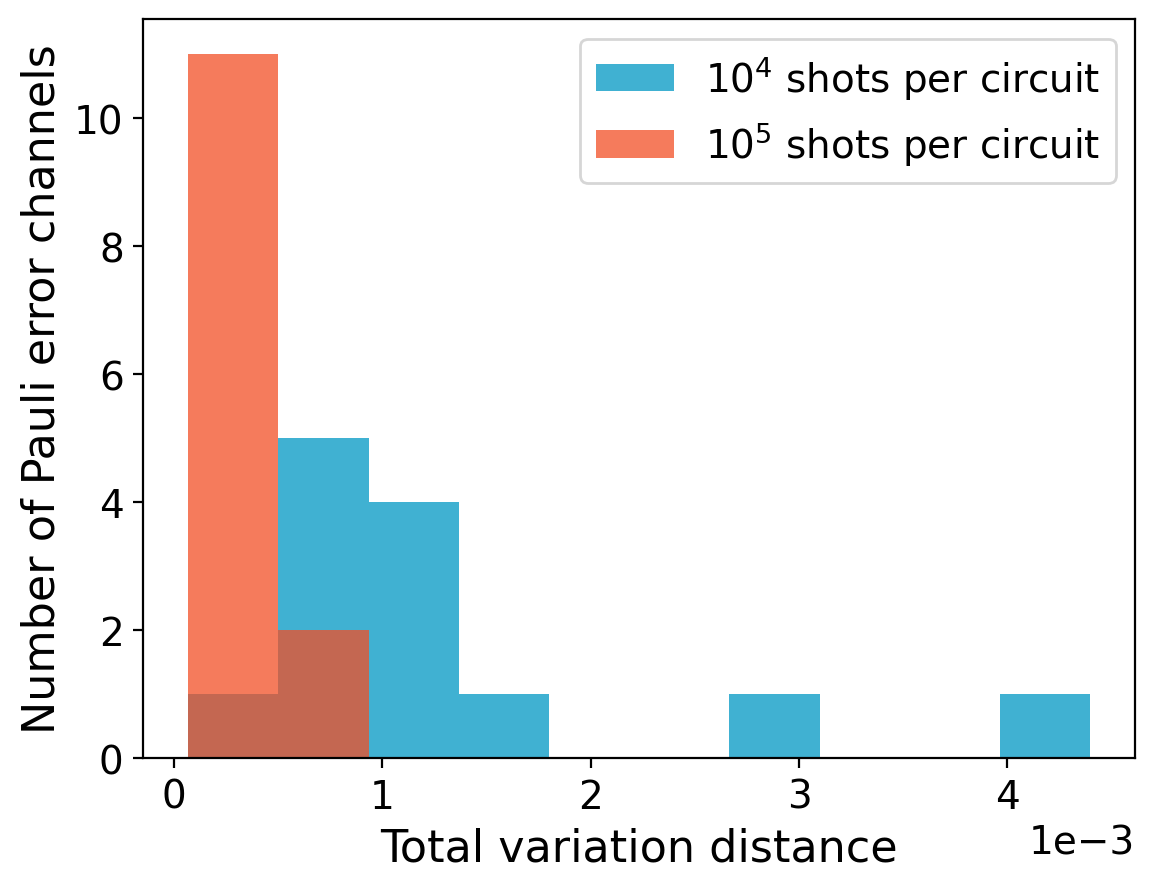}
    \caption{Total variation distance between estimated Pauli error rates and those used in simulation with $10^4$ and $10^5$ shots per circuit.}
    \label{fig:tvd-error-rates}
\end{figure}

We use the estimated gate eigenvalues to obtain estimated Pauli error rates. We measure the accuracy of this error rates using the total variation distance between discrete probability measures, defined as $\delta(P, Q) = \frac{1}{2} \sum_x \left|P(x) - Q(x) \right|$. The error rates for each gate are considered as a distribution, and we take the TVD between the error rates estimated with ACES and the rates used when simulating the circuits for each gate.
These values for our simulations are reported in Figure \ref{fig:tvd-error-rates}.

For the eigenvalue absolute error, we can see that increasing the number of shots by one magnitude makes the distribution get closer to $0$. We see the same behavior for the total variation distance. Further, note that the horizontal axis for both figures is given in scientific notation with magnitude indicated in the right-hand side. 


\subsection{Hardware}

We ran ACES on IBM Algiers and IBM Osaka. For both devices we only used a subset of their available qubits. Two qubits for Algiers and five (with linear connectivity) for Osaka. Note that, as mentioned before, we are not using twirling in these experimental runs. Although we see an advantage in using $10^5$ shots over $10^4$ from our simulations, we used the latter due to hardware constraints.

\begin{figure}
    \centering
    \includegraphics[scale=0.5]{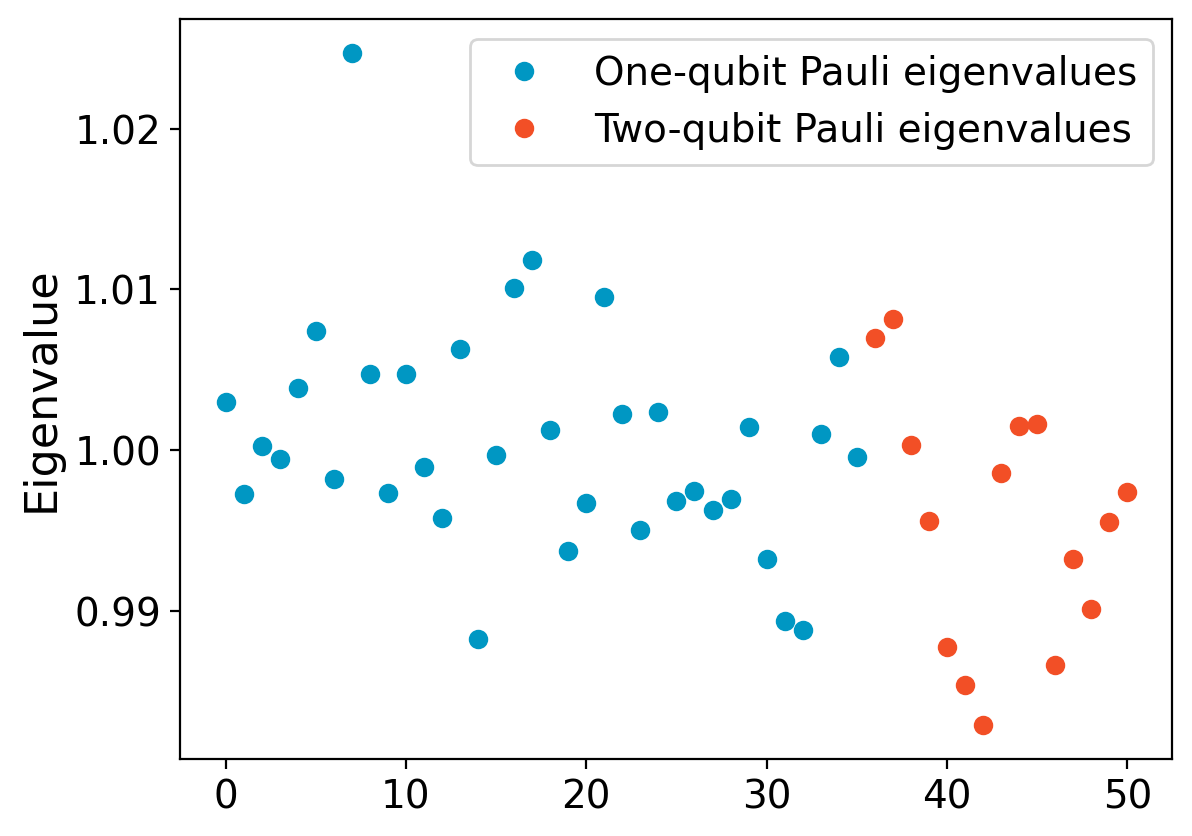}
    \includegraphics[scale=0.5]{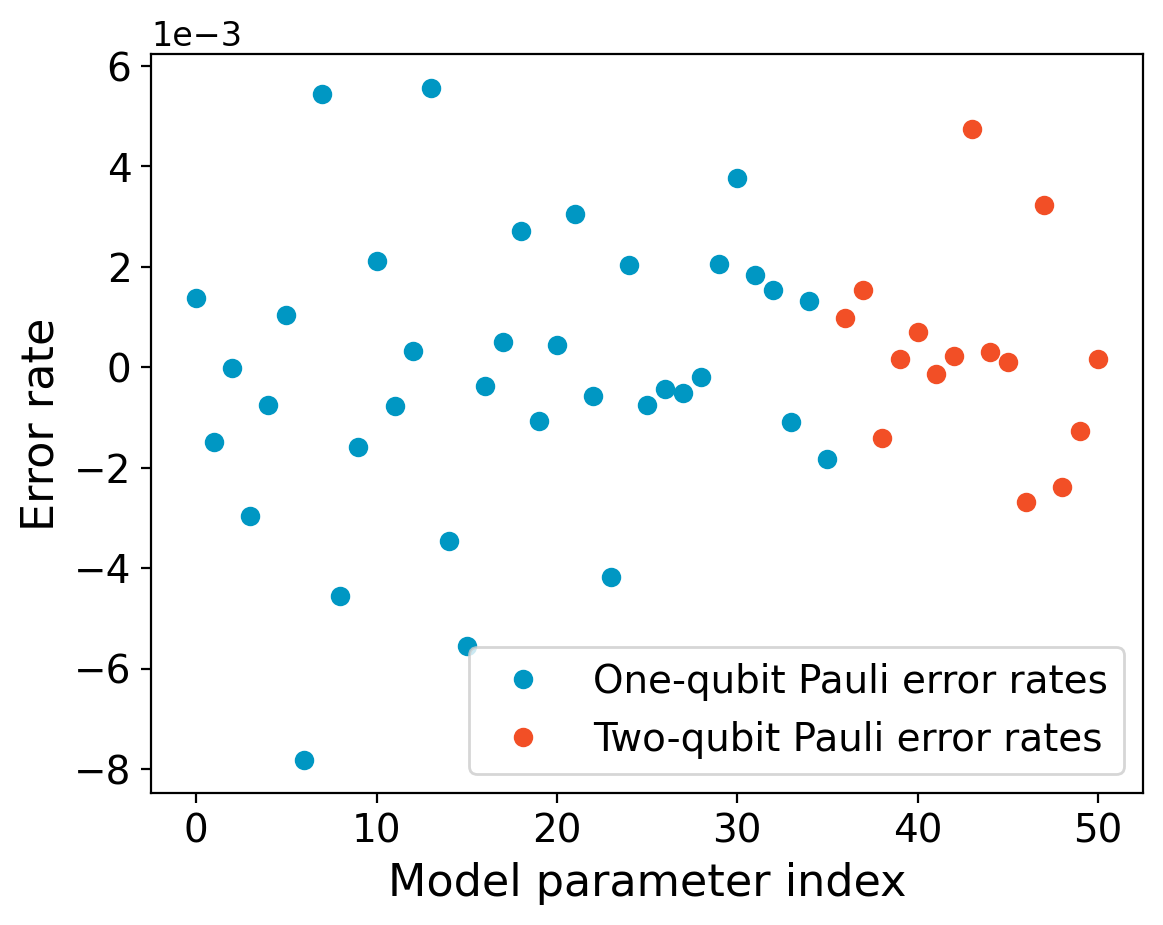}
    \caption{Estimated gate eigenvalues (top) and Pauli error rates (bottom) in qubits 0 and 1 of IBM Algiers.}
    \label{fig:algiers-results}
\end{figure}

\begin{figure}
    \centering
    \includegraphics[scale=0.5]{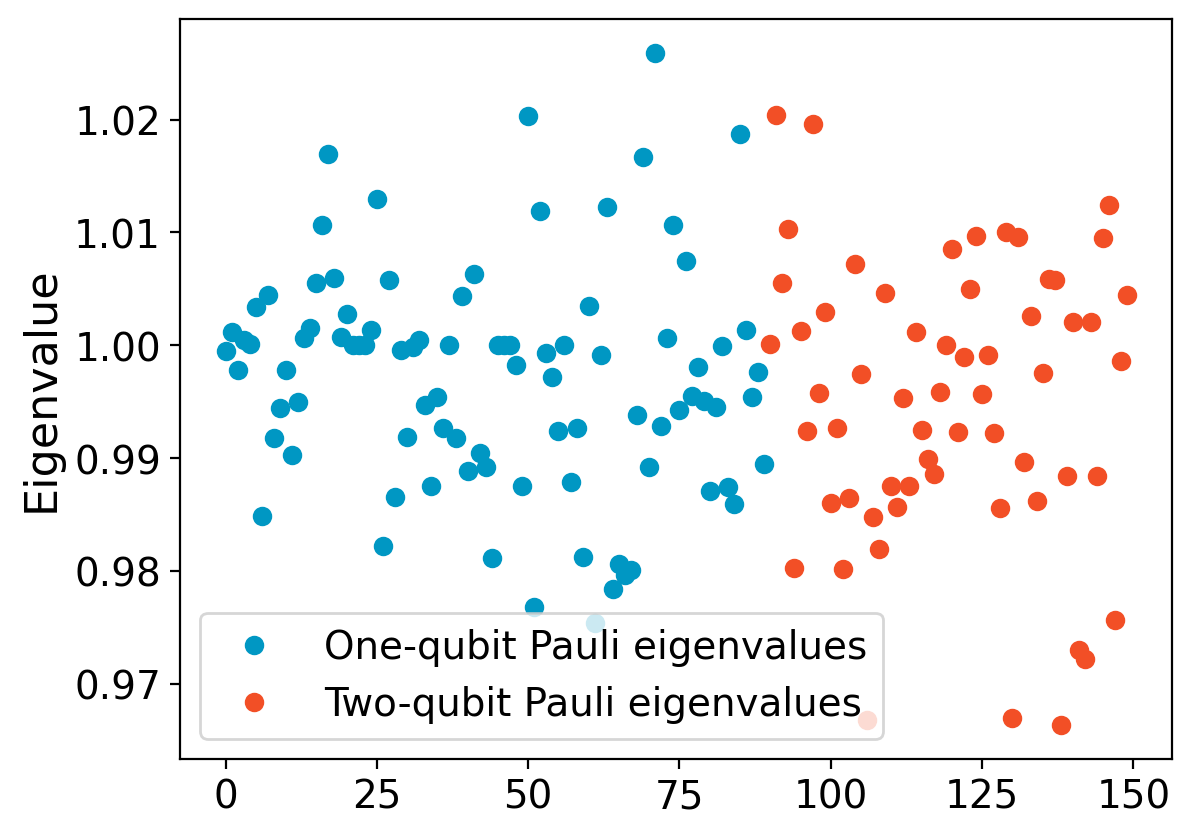}
    \includegraphics[scale=0.5]{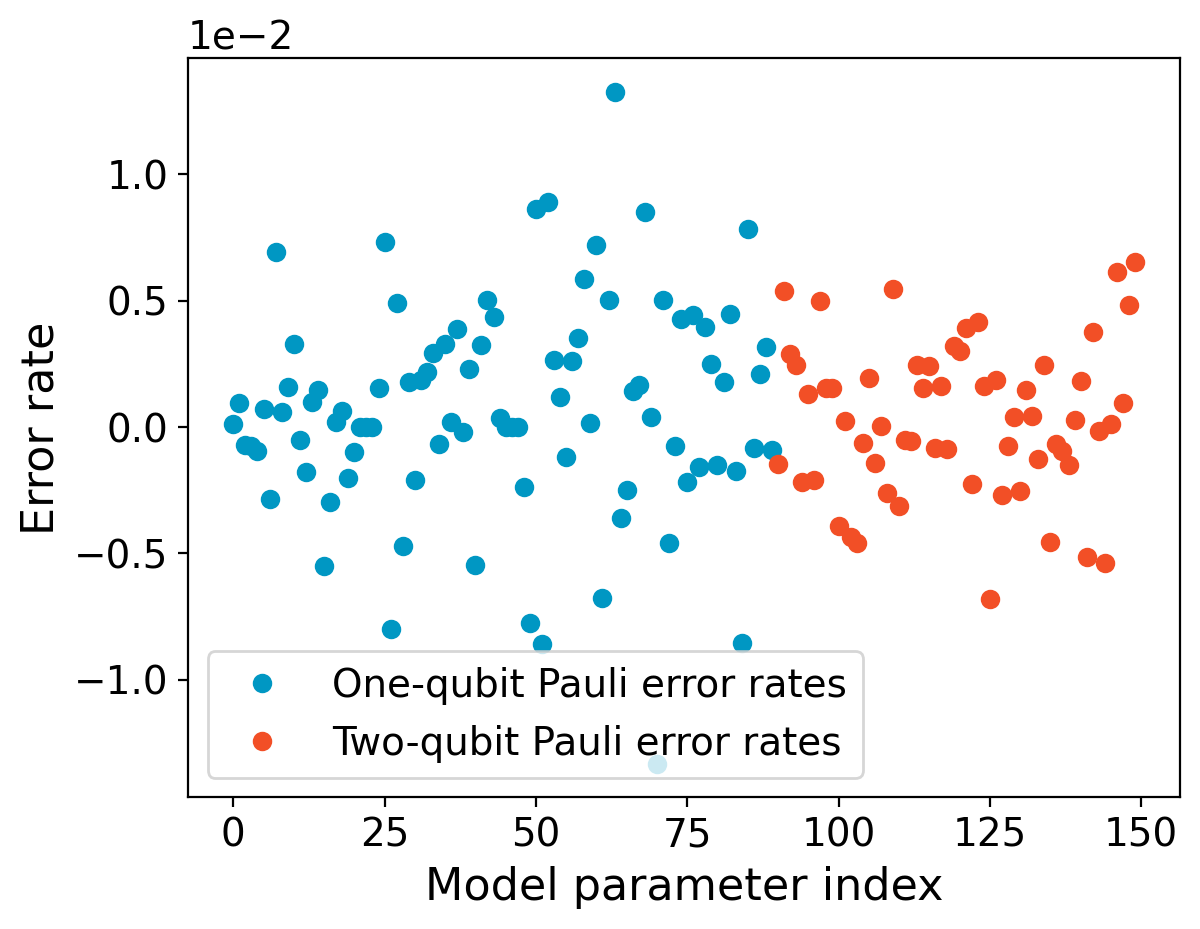}
    \caption{Estimated gate eigenvalues (top) and Pauli error rates (bottom) in qubits 0 through 4 of IBM Osaka.}
    \label{fig:osaka-results}
\end{figure}

In Figure \ref{fig:algiers-results}, we show the estimated eigenvalues and error rates from running ACES on qubits 0 and 1 of IBM Algiers. The horizontal axis ranges over the model parameter ($\upsilon$, as defined in Section \ref{sec:implementing}), starting with all the one-qubit gates for all qubits and then all the two-qubit gates for all two-qubit connections. Therefore, each point represents a single eigenvalue $\lambda_{\upsilon}$ (top) or single Pauli error rate $p_{\upsilon}$ (bottom) through \eqref{eq:eig-to-rates}. Similarly, the estimated eigenvalues and error rates from running ACES on qubits 0 through 4 of IBM Osaka are presented in Figure \ref{fig:osaka-results}. Note that for both figures, the vertical axis for the figures with estimated error rates is in scientific notation, with magnitude indicated at the top.

In both cases, we see that the eigenvalues are scattered around 1, and the error rates are scattered around 0. This is expected as the reported noise data for these devices indicates (see discussion below). 

\section{Discussion}\label{sec:discussion}

Through the simulation and hardware results, we find that ACES learns Pauli error channels successfully. In simulation, it is able to learn the individual Pauli error channel for each gate with an average total variation distance under $0.005$ using $10^4$ shots and under $0.001$ using $10^5$ shots. Similarly, the absolute value of the absolute error for the estimated individual gate eigenvalues is, on average, less than $0.01$ with $10^4$ shots and less than $0.003$ with $10^5$. In simulation, this was achieved only using $5$ circuits each with depth $14$. Although a high sampling rate is used in simulation, the low number of circuits and layers is promising. 

When interpreting the results from running the protocol in hardware, it is important to keep in mind that we are trying to estimate the hardware's noise channel as a stochastic Pauli channel. To estimate the success of the protocol in this scenario, we can use the reported two-gate fidelity by IBM. For Algiers, the calibration data at the time we ran our experiments indicated a $6.2\cdot10^{-3}$ CNOT error between qubits 0 and 1 of Algiers. Reconstructing the estimated noise channels, we get an infidelity of $7.6\cdot10^{-3}$ for the CZ gate between the same qubits. Reconstructing, for example, the error channel between qubits 3 and 4 in Osaka, we estimate an infidelity of $2.9\cdot10^{-2}$, while IBM reports a single-qubit ECR error for qubit 3 of $2.6\cdot10^{-2}$. However, this comparison should be taken with a grain of salt, as the following example illustrates. Looking at the CZ gate between qubits 0 and 1 for Osaka, we estimate an infidelity of $3.6\cdot10^{-2}$, while the reported single-qubit ECR error for qubit 2 is $3.8\cdot10^{-3}$, which is in disagreement by an order of magnitude. As we can see, in general, we get error rates greater than the values reported by IBM, which might be caused by not considering preparation and measurement errors separately, so that these are absorbed into the estimated gate error rates.

Next steps for this work include using the data learned about physical noise channels to perform error mitigation techniques to subsequently run circuits and correct for noise. This requires the subsequent circuits to run close in time to the characterizing protocol, as to keep noise conditions similar. An interesting question along this road is to explore the capabilities of both error estimation and mitigation for near-Clifford circuits with a small number of $T$ gates \cite{helsen2019new, smith2023cliffordbased}. On the same avenue, it is interesting to see how the protocol could be applied to any universal gateset. The protocol described in this paper should provide a good starting point for this. 

\section{Acknowledgements}

This material is supported by the U.S. Department of Energy, Office of Science, Office of Advanced Scientific Computing Research under Award Number DE-SC0021526. We acknowledge the use of IBM Quantum services for this work. The views expressed are those of the authors, and do not reflect the official policy or position of IBM or the IBM Quantum team. We thank the original author of the protocol, Steven Flammia, for helpful discussion. 

\bibliography{main}

\begin{thebibliography}{10}%
\makeatletter
\providecommand \@ifxundefined [1]{%
 \@ifx{#1\undefined}
}%
\providecommand \@ifnum [1]{%
 \ifnum #1\expandafter \@firstoftwo
 \else \expandafter \@secondoftwo
 \fi
}%
\providecommand \@ifx [1]{%
 \ifx #1\expandafter \@firstoftwo
 \else \expandafter \@secondoftwo
 \fi
}%
\providecommand \natexlab [1]{#1}%
\providecommand \enquote  [1]{``#1''}%
\providecommand \bibnamefont  [1]{#1}%
\providecommand \bibfnamefont [1]{#1}%
\providecommand \citenamefont [1]{#1}%
\providecommand \href@noop [0]{\@secondoftwo}%
\providecommand \href [0]{\begingroup \@sanitize@url \@href}%
\providecommand \@href[1]{\@@startlink{#1}\@@href}%
\providecommand \@@href[1]{\endgroup#1\@@endlink}%
\providecommand \@sanitize@url [0]{\catcode `\\12\catcode `\$12\catcode
  `\&12\catcode `\#12\catcode `\^12\catcode `\_12\catcode `\%12\relax}%
\providecommand \@@startlink[1]{}%
\providecommand \@@endlink[0]{}%
\providecommand \url  [0]{\begingroup\@sanitize@url \@url }%
\providecommand \@url [1]{\endgroup\@href {#1}{\urlprefix }}%
\providecommand \urlprefix  [0]{URL }%
\providecommand \Eprint [0]{\href }%
\providecommand \doibase [0]{https://doi.org/}%
\providecommand \selectlanguage [0]{\@gobble}%
\providecommand \bibinfo  [0]{\@secondoftwo}%
\providecommand \bibfield  [0]{\@secondoftwo}%
\providecommand \translation [1]{[#1]}%
\providecommand \BibitemOpen [0]{}%
\providecommand \bibitemStop [0]{}%
\providecommand \bibitemNoStop [0]{.\EOS\space}%
\providecommand \EOS [0]{\spacefactor3000\relax}%
\providecommand \BibitemShut  [1]{\csname bibitem#1\endcsname}%
\let\auto@bib@innerbib\@empty
\bibitem [{\citenamefont {Flammia}(2021)}]{Flammia2021AveragedCE}%
  \BibitemOpen
  \bibfield  {author} {\bibinfo {author} {\bibfnamefont {S.~T.}\ \bibnamefont
  {Flammia}},\ }\bibfield  {title} {\bibinfo {title} {Averaged circuit
  eigenvalue sampling},\ }in\ \href@noop {} {\emph {\bibinfo {booktitle}
  {Theory of Quantum Computation, Communication, and Cryptography}}}\ (\bibinfo
  {year} {2021})\BibitemShut {NoStop}%
\bibitem [{\citenamefont {Campbell}\ \emph {et~al.}(2023)\citenamefont
  {Campbell}, \citenamefont {Chong}, \citenamefont {Dahl}, \citenamefont
  {Frederick}, \citenamefont {Goiporia}, \citenamefont {Gokhale}, \citenamefont
  {Hall}, \citenamefont {Issa}, \citenamefont {Jones}, \citenamefont {Lee},
  \citenamefont {Litteken}, \citenamefont {Omole}, \citenamefont {Owusu-Antwi},
  \citenamefont {Perlin}, \citenamefont {Rines}, \citenamefont {Smith},
  \citenamefont {Goss}, \citenamefont {Hashim}, \citenamefont {Naik},
  \citenamefont {Younis}, \citenamefont {Lobser}, \citenamefont {Yale},
  \citenamefont {Huang},\ and\ \citenamefont {Liu}}]{superstaq}%
  \BibitemOpen
  \bibfield  {author} {\bibinfo {author} {\bibfnamefont {C.}~\bibnamefont
  {Campbell}}, \bibinfo {author} {\bibfnamefont {F.~T.}\ \bibnamefont {Chong}},
  \bibinfo {author} {\bibfnamefont {D.}~\bibnamefont {Dahl}}, \bibinfo {author}
  {\bibfnamefont {P.}~\bibnamefont {Frederick}}, \bibinfo {author}
  {\bibfnamefont {P.}~\bibnamefont {Goiporia}}, \bibinfo {author}
  {\bibfnamefont {P.}~\bibnamefont {Gokhale}}, \bibinfo {author} {\bibfnamefont
  {B.}~\bibnamefont {Hall}}, \bibinfo {author} {\bibfnamefont {S.}~\bibnamefont
  {Issa}}, \bibinfo {author} {\bibfnamefont {E.}~\bibnamefont {Jones}},
  \bibinfo {author} {\bibfnamefont {S.}~\bibnamefont {Lee}}, \bibinfo {author}
  {\bibfnamefont {A.}~\bibnamefont {Litteken}}, \bibinfo {author}
  {\bibfnamefont {V.}~\bibnamefont {Omole}}, \bibinfo {author} {\bibfnamefont
  {D.}~\bibnamefont {Owusu-Antwi}}, \bibinfo {author} {\bibfnamefont {M.~A.}\
  \bibnamefont {Perlin}}, \bibinfo {author} {\bibfnamefont {R.}~\bibnamefont
  {Rines}}, \bibinfo {author} {\bibfnamefont {K.~N.}\ \bibnamefont {Smith}},
  \bibinfo {author} {\bibfnamefont {N.}~\bibnamefont {Goss}}, \bibinfo {author}
  {\bibfnamefont {A.}~\bibnamefont {Hashim}}, \bibinfo {author} {\bibfnamefont
  {R.}~\bibnamefont {Naik}}, \bibinfo {author} {\bibfnamefont {E.}~\bibnamefont
  {Younis}}, \bibinfo {author} {\bibfnamefont {D.}~\bibnamefont {Lobser}},
  \bibinfo {author} {\bibfnamefont {C.~G.}\ \bibnamefont {Yale}}, \bibinfo
  {author} {\bibfnamefont {B.}~\bibnamefont {Huang}},\ and\ \bibinfo {author}
  {\bibfnamefont {J.}~\bibnamefont {Liu}},\ }\bibfield  {title} {\bibinfo
  {title} {{Superstaq: Deep Optimization of Quantum Programs}},\ }in\
  \href@noop {} {\emph {\bibinfo {booktitle} {{2023 International Conference on
  Quantum Computing and Engineering}}}}\ (\bibinfo {year} {2023})\ \Eprint
  {https://arxiv.org/abs/2309.05157} {arXiv:2309.05157 [quant-ph]} \BibitemShut
  {NoStop}%
\bibitem [{\citenamefont {Cao}\ \emph {et~al.}(2022)\citenamefont {Cao},
  \citenamefont {Lin}, \citenamefont {Kribs}, \citenamefont {Poon},
  \citenamefont {Zeng},\ and\ \citenamefont {Laflamme}}]{cao2022nisq}%
  \BibitemOpen
  \bibfield  {author} {\bibinfo {author} {\bibfnamefont {N.}~\bibnamefont
  {Cao}}, \bibinfo {author} {\bibfnamefont {J.}~\bibnamefont {Lin}}, \bibinfo
  {author} {\bibfnamefont {D.}~\bibnamefont {Kribs}}, \bibinfo {author}
  {\bibfnamefont {Y.-T.}\ \bibnamefont {Poon}}, \bibinfo {author}
  {\bibfnamefont {B.}~\bibnamefont {Zeng}},\ and\ \bibinfo {author}
  {\bibfnamefont {R.}~\bibnamefont {Laflamme}},\ }\href@noop {} {\bibinfo
  {title} {Nisq: Error correction, mitigation, and noise simulation}} (\bibinfo
  {year} {2022}),\ \Eprint {https://arxiv.org/abs/2111.02345} {arXiv:2111.02345
  [quant-ph]} \BibitemShut {NoStop}%
\bibitem [{\citenamefont {Emerson}\ \emph {et~al.}(2005)\citenamefont
  {Emerson}, \citenamefont {Alicki},\ and\ \citenamefont
  {{\.Z}yczkowski}}]{emerson2005scalable}%
  \BibitemOpen
  \bibfield  {author} {\bibinfo {author} {\bibfnamefont {J.}~\bibnamefont
  {Emerson}}, \bibinfo {author} {\bibfnamefont {R.}~\bibnamefont {Alicki}},\
  and\ \bibinfo {author} {\bibfnamefont {K.}~\bibnamefont {{\.Z}yczkowski}},\
  }\bibfield  {title} {\bibinfo {title} {Scalable noise estimation with random
  unitary operators},\ }\href@noop {} {\bibfield  {journal} {\bibinfo
  {journal} {Journal of Optics B: Quantum and Semiclassical Optics}\ }\textbf
  {\bibinfo {volume} {7}},\ \bibinfo {pages} {S347} (\bibinfo {year}
  {2005})}\BibitemShut {NoStop}%
\bibitem [{\citenamefont {Knill}\ \emph {et~al.}(2008)\citenamefont {Knill},
  \citenamefont {Leibfried}, \citenamefont {Reichle}, \citenamefont {Britton},
  \citenamefont {Blakestad}, \citenamefont {Jost}, \citenamefont {Langer},
  \citenamefont {Ozeri}, \citenamefont {Seidelin},\ and\ \citenamefont
  {Wineland}}]{Knill_2008}%
  \BibitemOpen
  \bibfield  {author} {\bibinfo {author} {\bibfnamefont {E.}~\bibnamefont
  {Knill}}, \bibinfo {author} {\bibfnamefont {D.}~\bibnamefont {Leibfried}},
  \bibinfo {author} {\bibfnamefont {R.}~\bibnamefont {Reichle}}, \bibinfo
  {author} {\bibfnamefont {J.}~\bibnamefont {Britton}}, \bibinfo {author}
  {\bibfnamefont {R.~B.}\ \bibnamefont {Blakestad}}, \bibinfo {author}
  {\bibfnamefont {J.~D.}\ \bibnamefont {Jost}}, \bibinfo {author}
  {\bibfnamefont {C.}~\bibnamefont {Langer}}, \bibinfo {author} {\bibfnamefont
  {R.}~\bibnamefont {Ozeri}}, \bibinfo {author} {\bibfnamefont
  {S.}~\bibnamefont {Seidelin}},\ and\ \bibinfo {author} {\bibfnamefont
  {D.~J.}\ \bibnamefont {Wineland}},\ }\bibfield  {title} {\bibinfo {title}
  {Randomized benchmarking of quantum gates},\ }\bibfield  {journal} {\bibinfo
  {journal} {Physical Review A}\ }\textbf {\bibinfo {volume} {77}},\ \href
  {https://doi.org/10.1103/physreva.77.012307} {10.1103/physreva.77.012307}
  (\bibinfo {year} {2008})\BibitemShut {NoStop}%
\bibitem [{\citenamefont {Nielsen}(2002)}]{Nielsen_2002}%
  \BibitemOpen
  \bibfield  {author} {\bibinfo {author} {\bibfnamefont {M.~A.}\ \bibnamefont
  {Nielsen}},\ }\bibfield  {title} {\bibinfo {title} {A simple formula for the
  average gate fidelity of a quantum dynamical operation},\ }\href
  {https://doi.org/10.1016/s0375-9601(02)01272-0} {\bibfield  {journal}
  {\bibinfo  {journal} {Physics Letters A}\ }\textbf {\bibinfo {volume}
  {303}},\ \bibinfo {pages} {249–252} (\bibinfo {year} {2002})}\BibitemShut
  {NoStop}%
\bibitem [{\citenamefont {Wallman}\ and\ \citenamefont
  {Emerson}(2016)}]{Wallman_2016}%
  \BibitemOpen
  \bibfield  {author} {\bibinfo {author} {\bibfnamefont {J.~J.}\ \bibnamefont
  {Wallman}}\ and\ \bibinfo {author} {\bibfnamefont {J.}~\bibnamefont
  {Emerson}},\ }\bibfield  {title} {\bibinfo {title} {Noise tailoring for
  scalable quantum computation via randomized compiling},\ }\bibfield
  {journal} {\bibinfo  {journal} {Physical Review A}\ }\textbf {\bibinfo
  {volume} {94}},\ \href {https://doi.org/10.1103/physreva.94.052325}
  {10.1103/physreva.94.052325} (\bibinfo {year} {2016})\BibitemShut {NoStop}%
\bibitem [{\citenamefont {Hashim}\ \emph {et~al.}(2021)\citenamefont {Hashim},
  \citenamefont {Naik}, \citenamefont {Morvan}, \citenamefont {Ville},
  \citenamefont {Mitchell}, \citenamefont {Kreikebaum}, \citenamefont {Davis},
  \citenamefont {Smith}, \citenamefont {Iancu}, \citenamefont {O'Brien},
  \citenamefont {Hincks}, \citenamefont {Wallman}, \citenamefont {Emerson},\
  and\ \citenamefont {Siddiqi}}]{hashim2021randomized}%
  \BibitemOpen
  \bibfield  {author} {\bibinfo {author} {\bibfnamefont {A.}~\bibnamefont
  {Hashim}}, \bibinfo {author} {\bibfnamefont {R.~K.}\ \bibnamefont {Naik}},
  \bibinfo {author} {\bibfnamefont {A.}~\bibnamefont {Morvan}}, \bibinfo
  {author} {\bibfnamefont {J.-L.}\ \bibnamefont {Ville}}, \bibinfo {author}
  {\bibfnamefont {B.}~\bibnamefont {Mitchell}}, \bibinfo {author}
  {\bibfnamefont {J.~M.}\ \bibnamefont {Kreikebaum}}, \bibinfo {author}
  {\bibfnamefont {M.}~\bibnamefont {Davis}}, \bibinfo {author} {\bibfnamefont
  {E.}~\bibnamefont {Smith}}, \bibinfo {author} {\bibfnamefont
  {C.}~\bibnamefont {Iancu}}, \bibinfo {author} {\bibfnamefont {K.~P.}\
  \bibnamefont {O'Brien}}, \bibinfo {author} {\bibfnamefont {I.}~\bibnamefont
  {Hincks}}, \bibinfo {author} {\bibfnamefont {J.~J.}\ \bibnamefont {Wallman}},
  \bibinfo {author} {\bibfnamefont {J.}~\bibnamefont {Emerson}},\ and\ \bibinfo
  {author} {\bibfnamefont {I.}~\bibnamefont {Siddiqi}},\ }\bibfield  {title}
  {\bibinfo {title} {Randomized compiling for scalable quantum computing on a
  noisy superconducting quantum processor},\ }\href
  {https://doi.org/10.1103/PhysRevX.11.041039} {\bibfield  {journal} {\bibinfo
  {journal} {Phys. Rev. X}\ }\textbf {\bibinfo {volume} {11}},\ \bibinfo
  {pages} {041039} (\bibinfo {year} {2021})}\BibitemShut {NoStop}%
\bibitem [{\citenamefont {Helsen}\ \emph {et~al.}(2019)\citenamefont {Helsen},
  \citenamefont {Xue}, \citenamefont {Vandersypen},\ and\ \citenamefont
  {Wehner}}]{helsen2019new}%
  \BibitemOpen
  \bibfield  {author} {\bibinfo {author} {\bibfnamefont {J.}~\bibnamefont
  {Helsen}}, \bibinfo {author} {\bibfnamefont {X.}~\bibnamefont {Xue}},
  \bibinfo {author} {\bibfnamefont {L.~M.~K.}\ \bibnamefont {Vandersypen}},\
  and\ \bibinfo {author} {\bibfnamefont {S.}~\bibnamefont {Wehner}},\
  }\href@noop {} {\bibinfo {title} {A new class of efficient randomized
  benchmarking protocols}} (\bibinfo {year} {2019}),\ \Eprint
  {https://arxiv.org/abs/1806.02048} {arXiv:1806.02048 [quant-ph]} \BibitemShut
  {NoStop}%
\bibitem [{\citenamefont {Smith}\ \emph {et~al.}(2023)\citenamefont {Smith},
  \citenamefont {Perlin}, \citenamefont {Gokhale}, \citenamefont {Frederick},
  \citenamefont {Owusu-Antwi}, \citenamefont {Rines}, \citenamefont {Omole},\
  and\ \citenamefont {Chong}}]{smith2023cliffordbased}%
  \BibitemOpen
  \bibfield  {author} {\bibinfo {author} {\bibfnamefont {K.~N.}\ \bibnamefont
  {Smith}}, \bibinfo {author} {\bibfnamefont {M.~A.}\ \bibnamefont {Perlin}},
  \bibinfo {author} {\bibfnamefont {P.}~\bibnamefont {Gokhale}}, \bibinfo
  {author} {\bibfnamefont {P.}~\bibnamefont {Frederick}}, \bibinfo {author}
  {\bibfnamefont {D.}~\bibnamefont {Owusu-Antwi}}, \bibinfo {author}
  {\bibfnamefont {R.}~\bibnamefont {Rines}}, \bibinfo {author} {\bibfnamefont
  {V.}~\bibnamefont {Omole}},\ and\ \bibinfo {author} {\bibfnamefont {F.~T.}\
  \bibnamefont {Chong}},\ }\href@noop {} {\bibinfo {title} {Clifford-based
  circuit cutting for quantum simulation}} (\bibinfo {year} {2023}),\ \Eprint
  {https://arxiv.org/abs/2303.10788} {arXiv:2303.10788 [quant-ph]} \BibitemShut
  {NoStop}%
\end{thebibliography}%

\end{document}